\documentclass[12pt]{article}
\input xy
\xyoption{all}
\usepackage{amsmath,epsf,amssymb}
\newcommand{\f} {\mathcal{F}}
\newcommand{\la}{\longrightarrow}

\newcommand{\x}{\mathsf{x}}
\newcommand{\y}{\mathsf{y}}
\renewcommand{\k}{\mathsf{K}}
\renewcommand{\j}{\mathsf{J}}
\newcommand{\F}{\mathsf{F}}
\renewcommand{\o}{\mathcal{O}}

\begin{document}

\hfill $\,$

\vspace{1.0in}

\begin{center}

{\large\bf D-branes, obstructed curves, and minimal model superpotentials}

\vspace{0.5in}

Gueorgui Todorov \\
Department of Mathematics\\
University of Utah\\
Salt Lake City, UT  84112 \\
{\tt todorov@math.utah.edu} \\

$\,$

\end{center}

In this short note we apply methods of Aspinwall-Katz to compute
superpotentials of D-branes wrapped on more general obstructed rational
curves in Calabi-Yau threefolds.  We find an a priori unexpected match
between superpotentials from certain such curves and the superpotentials
of Landau-Ginzburg models corresponding to minimal models.

\begin{flushleft}
September 2007
\end{flushleft}

\newpage

\tableofcontents

\section{Introduction}

The application of derived categories to D-branes in physics,
originally described in \cite{esdc} and later popularized in
\cite{mikedc} (see \cite{eslec} for a review), 
has proven to be a very important techical tool
in mathematical string theory.
This program has yielded results ranging from new notions of
stability \cite{bridgeland1,bridgeland2} to, most recently, the construction
of CFT's for Kontsevich's nc spaces \cite{cdhps}, which are defined
in terms of their sheaf theory. 

Part of the reason that the derived categories program has been so useful
is that in principle it gives a complete understanding of the
off-shell states in the open string B model, meaning that in principle
not only can one directly compute all massless spectra of open strings,
but also all correlation functions between massless states.

The first direct computation of massless spectra of open strings
between D-branes on subvarieties of the target space appeared
in \cite{ks}, where, after taking into account the Minasian-Moore-Freed-Witten
anomaly \cite{rubengreg,freeded} and the open string B model 
anomaly, it was shown, for example, 
that the worldsheet 
CFT computation
realizes a spectral sequence.

Ideally, one would like to next directly compute couplings from
those massless spectrum computations.  In the case of the
closed string B model, this is fairly trivial, but in the open
string B model, this is rather more complicated.
We shall outline a direct computation of massless spectra of open
strings beginning and ending on a D-brane wrapped on an obstructed
curve in section~\ref{direct}, and as we shall see there,
the computation implies a connection between curvatures of
Riemannian metrics and obstructions in deformation theory which
we have not yet been able to verify.

However, there are other approaches to such problems.
The point of the derived-categories-in-physics program
\cite{esdc,mikedc,eslec}
is that derived categories classify universality classes of
open string boundary states, so, computations that are difficult
with some representatives may be replaced with other computations
involving different representatives of the same universality class.
By replacing D-branes wrapped on obstructed curves
with brane/antibrane systems in the same
universality class, with each brane and antibrane covering the
entire space, one gets a much more nearly straightforward
computation.
This method was used in \cite{aspkatz} to describe how to
compute all couplings between open string B model states,
reproducing the full $A_{\infty}$ algebra structure of open string
field theory \cite{zwiebach}.

Of course, the drawback of this method is the same intrinsic
to all work in the derived-categories-in-physics program:
we do not know for certain that physical universality classes really
do coincide with equivalence classes in the derived category.
Numerous tests of this conjecture have been performed by various
authors, so it is widely believed to be true,
but as a matter of principle, there is a fundamental issue here.
(There is an analogous issue that arises when discussing stacks in
physics \cite{ps1,ps2,ps3,ps4}.  There, the issue is that a given stack 
has several different
presentations which can be very different QFT's; the relevant
conjecture is that universality classes are classified by stacks.
This, also, has now been checked in numerous different ways.)
See \cite{ps5} for an overview of such connections between
universality classes in physics and mathematical equivalences.

In section~\ref{ak} we shall use the methods of \cite{aspkatz} to compute
couplings / superpotential terms from D-branes wrapped on 
obstructed curves appearing in small resolutions,
the same issue which we attempted via a direct computation
in section~\ref{direct}.
Curiously, we will find that D-branes wrapped on obstructed
curves in ADE-type three-folds possess the same superpotentials
as ADE-type minimal models.
To be precise, recall that minimal models in two-dimensional
CFT's have Landau-Ginzburg descriptions with an ADE classification,
summarized in the table below \cite{vafawarner}:
\begin{center}
\begin{tabular}{cc}
Algebra & Superpotential \\ \hline
$A_n, n \geq 1$ & $x^{n+1}$ \\
$D_n, n \geq 1$ & $x^{n-1} + x y^2$ \\
$E_6$ & $x^3 + y^4$ \\
$E_7$ & $x^3 + xy^3$ \\
$E_8$ & $x_3 + y^5$
\end{tabular}
\end{center}
These superpotentials will be reproduced by D-branes wrapped on
obstructed ${\bf P}^1$'s in Calabi-Yau threefolds.  In such cases, the normal
bundle will have one of the following three forms:
\begin{itemize}
\item ${\cal O}(-1) \oplus {\cal O}(-1)$
\item ${\cal O} \oplus {\cal O}(-2)$
\item ${\cal O}(1) \oplus {\cal O}(-3)$
\end{itemize}
The first case has no infinitesimal deformations, and so is uninteresting
for our purposes in this paper.  The second case has a single
(obstructed) infinitesimal deformation, and this case will give rise
to superpotentials of the $A_n$ form, where the field $x$ corresponds
to that one infinitesimal deformation.  The third case has
two (obstructed) infinitesimal deformations, and this case will give
the $D_n$ and $E_n$ series, with the $x$ and $y$ fields corresponding
to those two infinitesimal deformations.
See for example \cite{laufer,shel-dave} for more information 
on these geometries,
which are small resolutions of singular Calabi-Yau threefolds.

Superpotentials for D-branes wrapped on obstructed curves
have been of interest in many other places in the physics
literature.  For example, such wrapped D-branes made an appearance
in \cite{dv}, where they were used to motivate a gauge theory
having an adjoint-valued field $\phi$ with a $\phi^n$-type
superpotential\footnote{The superpotential was checked indirectly in
\cite[section 2.2]{shamitkatz}
using a dimensionally-reduced holomorphic Chern-Simons theory
(implicitly assuming the dimensional reduction of the open string field
theory on the total space coincides with open string field theory of
D-branes on a submanifold).  However, a direct derivation in open string
CFT was
not given in that paper.}.

Related work has appeared in \cite{curto1,curto2}, where given superpotentials
of the form we consider, corresponding singular Calabi-Yau threefolds
were constructed.  In essence, that work considered the problem
inverse to that in this paper, by constructing geometry from superpotentials
instead of superpotentials from geometry.

\section{Outline of a direct computation}   \label{direct}

So that the reader will better appreciate the computational efficiency
of the derived categories program and
the methods of \cite{aspkatz}, in this section we will outline how
one could attempt a direct physical computation of superpotentials from
D-branes wrapped on obstructed curves.
This section will closely\footnote{We would like to thank E.~Sharpe
for giving us permission to reproduce his argument here.} 
follow \cite{eslec}[section 11.1].

We shall consider a single D-brane wrapped on an obstructed ${\bf P}^1$,
which is to say, a ${\bf P}^1$ which admits an infinitesimal deformation,
but whose deformation is obstructed at some order.

We shall assume that the gauge bundle on the D-brane is trivial,
so the boundary conditions on worldsheet fields take a simple form.
Furthermore, the restriction of the tangent bundle of the Calabi-Yau
to the ${\bf P}^1$ does split holomorphically.  Thus, neither of the usual
subtleties associated with open string computations is relevant here.

The normal bundle to the ${\bf P}^1$ is ${\cal O} \oplus {\cal O}(-2)$.
Since the normal bundle admits a holomorphic section, 
$\mbox{Ext}^1\left( {\cal O}_{ {\bf P}^1 }, {\cal O}_{ {\bf P}_1 } \right)$
is one-dimensional.  
For `generic' obstructions ({\it i.e.} of order 3),
the Yoneda pairing
\begin{displaymath}
\mbox{Ext}^1\left( {\cal O}_{ {\bf P}^1 }, {\cal O}_{ {\bf P}_1 } \right)
\times
\mbox{Ext}^1\left( {\cal O}_{ {\bf P}^1 }, {\cal O}_{ {\bf P}_1 } \right)
 \: \longrightarrow \:
\mbox{Ext}^2\left( {\cal O}_{ {\bf P}^1 }, {\cal O}_{ {\bf P}_1 } \right)
\end{displaymath}
is nonzero, and the obstruction is characterized by the image in Ext$^2$.
For (nongeneric) obstructions of higher order,
the Yoneda pairing will vanish, but a higher-order computation will
be nonvanishing.

Already at the level of vertex operators we can begin to see
some of the complications involved in realizing the Yoneda pairing.
In the present example, both Ext$^1$ and Ext$^2$ above are one-dimensional.
In fact,
\begin{eqnarray*}
\mbox{Ext}^1\left( {\cal O}_{ {\bf P}^1 }, {\cal O}_{ {\bf P}_1 } \right)
& = & H^0\left( {\cal N}_{ {\bf P}^1/X } \right) \: = \: {\bf C} \\
\mbox{Ext}^2\left( {\cal O}_{ {\bf P}^1 }, {\cal O}_{ {\bf P}_1 } \right)
& = & H^1\left( {\cal N}_{ {\bf P}^1/X } \right) \: = \: {\bf C} 
\end{eqnarray*}
From our earlier description of vertex operators, and the fact that
the only holomorphic section of ${\cal O}$ is the constant section,
we see that the elements of Ext$^1$ are described by the vertex operator
$\theta$ (associated to the ${\cal O}$ factor in the normal bundle),
and elements of Ext$^2$ are described by the vertex operator $\eta \theta$.
If the Yoneda pairing in this case were as trivial as just a wedge product,
then the image in Ext$^2$ would just be a product of $\theta$'s
 -- but by the Grassman property, such a product vanishes.
Instead, in a case in which the Yoneda pairing is nontrivial,
the image in Ext$^2$ is $\eta \theta$ instead of $\theta \theta$
 -- so the operator product must necessarily involve some sort of
interaction term that has the effect of changing a $\theta$ into
an $\eta$.

The fact that the normal bundle has this form might confuse the reader
 -- after all, the ${\bf P}^1$ is supposed to be obstructed,
and yet there is a one-parameter-family of rational curves inside the
normal bundle containing the ${\bf P}^1$.
The solution to this puzzle gives another reason why the Yoneda pairing
computation in this case is extremely difficult.
Unlike differential geometry, where normal bundles capture
local geometry, in algebraic geometry the normal bundle need {\it not}
encode the local holomorphic structure, only the local smooth structure.
In the present case, local coordinates in a neighborhood of the
obstructed ${\bf P}^1$ can be described as follows.
Let one coordinate patch on a holomorphic neighborhood have coordinates
$(x, y_1, y_2)$, and the other coordinate patch on a holomorphic neighborhood
have coordinates $(w, z_1, z_2)$, where
\begin{eqnarray*}
w & = & x^{-1} \\
z_1 & = & x^2 y_1 \: + \: x y_2^n \\
z_2 & = & y_2
\end{eqnarray*}
The integer $n$ is the degree of the obstruction,
the coordinates $x$, $w$ are coordinates on the ${\bf P}^1$, 
$z_2 = y_2$ is a coordinate on the ${\cal O}$ factor on the normal bundle,
and $z_1$, $y_1$ morally would be coordinates on the ${\cal O}(-2)$ factor,
except that the coordinate transformation is {\it not} that of
${\cal O}(-2)$ -- it's complicated by the $x y_2^n$ term, which means that
this local holomorphic neighborhood is not equivalent to the normal bundle.
The normal bundle is only a linearized approximation to local holomorphic
coordinates.  Unfortunately, data concerning the degree of the obstruction
({\it i.e.} the `extra' term in the expression for $z_1$)
is omitted by the linearization that gives rise to the normal bundle.

Thus, in order to see the obstruction, we need more data than the normal
bundle itself provides.  In order to recover the obstruction, the
BCFT calculation corresponding to the Yoneda pairing must have some
nonlocal component.

So, already before trying to set up the physics calculation, we see
two features that the result must have:
\begin{itemize}
\item The calculation must take advantage of some interaction term
in the worldsheet action -- the result is not just a wedge product,
unlike the closed string B model bulk-bulk OPE's.
\item The calculation must give a result that is somehow nonlocal.
\end{itemize}

Next, let us perform the calculation.
In principle, for a generic (order 3) obstruction, the following
three-point correlation function should vanish:
\begin{displaymath}
< \theta(\tau_1) \theta(\tau_2) \theta(\tau_3) >
\end{displaymath}
involving vertex operators for three copies of the element of Ext$^1$
inserted at various places along the boundary.
This correlation function should encode the Yoneda pairing, as outlined
earlier.

Now, in topological field theories, correlation functions should
reduce to zero modes.  In the present case, there is one $\eta$
zero mode and two $\theta$ zero modes, yet here we have three $\theta$'s.
The only way to get a nonzero result is to use some interaction terms.

Put another way, this correlation function should encode three
copies of the Yoneda pairing -- one for each pair of $\theta$'s.
In principle, each boundary-boundary OPE should take two $\theta$'s
and generate a $\eta \theta$ term, so that the result is a correlation
function involving one $\eta$ and two $\theta$'s, perfect to match
the available zero modes.  However, in order for the OPE to operate
in this fashion, we shall need some sort of interaction term.

Ordinarily one available interaction term would be the boundary interaction
\begin{displaymath}
\int_{\partial \Sigma} F_{i \overline{\jmath}} \rho^i \eta^{\overline{\jmath}}
\end{displaymath}
We could contract the $\rho$ on one of the $\theta$'s,
leaving us with two $\theta$'s and one $\eta$, perfect to match the
available zero modes.  The $\rho-\theta$ contraction
would generate a propagator factor proportional to $1/z$,
and the boundary integral would give a scale-invariant result.
The obvious log divergence can be handled by regularizing the propagator,
as discussed in \cite{edcs}, leaving a factor of an inverted laplacian.

In the present case, the curvature of the Chan-Paton factors can
be assumed to be trivial, so there is no such available interaction term,
but the general idea is on the right track.

The only available interaction term is the bulk four-fermi term:
\begin{displaymath}
\int_{\Sigma} R_{i \overline{\imath} j \overline{\jmath}}
\rho^i \rho^j \eta^{ \overline{\imath} } g^{\overline{\jmath} k}
\theta_k 
\end{displaymath}
We could contract the two $\rho$'s on two of the three $\theta$'s,
leaving us with a total of two $\theta$'s (one from the interaction term,
plus one of the original correlators) and one $\eta$, exactly as needed
to match the available zero modes.  Each $\rho-\theta$ contraction
would generate a propagator factor proportional to $1/z$, 
which would be cancelled by the integral over the bulk of the disk.
Boundary divergences can be handled by regularizing the propagators,
leaving us with factors of inverted laplacians.

Thus, we see the structure that we predicted earlier -- the correlation
function is nonvanishing thanks to an interaction term, and we have
nonlocal effects due to the presence of inverted laplacians.

What remains is to check that the resulting expression really does
correctly calculate the Yoneda pairing, which has not yet been
completed.

\section{Application of Aspinwall-Katz's methods} \label{ak}

In this section we will explicitly describe some examples of
superpotentials from wrapped branes using the methods of 
Aspinwall-Katz \cite{aspkatz}.  
As anticipated elsewhere, the resulting superpotentials
have the same form as in minimal models, yielding another connection
between geometry and physics.

The wrapped D-brane superpotentials are determined by an
$A_{\infty}$ structure. 
Following \cite{aspkatz}, the basic idea is that we will compute
the $A_{\infty}$ structure
encoded in D-brane superpotentials by replacing the original sheaves
modelling the wrapped D-branes with a different representative in the
derived category, one for which $A_{\infty}$ computations are much easier,
then compute the $A_{\infty}$ structures using those alternate
representatives.  In particular, this computation is much easier and
far more general than the attempted computation in the previous
section.
This also shows how the application of derived categories
to physics yields powerful technical tools.

In more detail, we are going to compute the $A_\infty$ structure as follows.
First, replace each object 
in in the derived category with a
quasi-isomorphic complex of injective sheaves. We may view this
as an injective resolution of these objects. 
 Suppose  for simplicity that we
have only one D-brane $\mathcal{E}^\bullet$. Then  the complex  of interest is 
with entries $ \oplus_p \textrm{Hom}(\mathcal{E}^p,\mathcal{E}^{p+n})$.
 If we denote an element of this group by $\sum_p f_{n,p}$,  
then the differential for this complex is
given by $\partial_n f_{n,p} = \textrm{d}_{p+n}\circ f_{n,p} - 
    (-1)^n f_{p+1,n}\circ \textrm{d}_{p}$ (cf. \cite[Equation (66)]{aspkatz}).
The composition gives this complex a dga structure.
But now by a Theorem of Kadeishvili \cite{kad} there is an $A_\infty$ 
structure on the cohomology of this complex with differential zero 
(minimal) and an $A_\infty$-morphism such that the first level is a 
chosen embedding of the cohomology in the complex. 
In our case this embedding will be very natural.
 This $A_\infty$ structure is not unique but it is unique up 
to $A_\infty$-isomorphism\footnote{In general, it does not seem that
all $A_{\infty}$ isomorphisms preserve the kinetic terms of the field theory,
so, strictly speaking, we are only interested in a subset of all
$A_{\infty}$ isomorphisms.  In addition, there is an issue that to describe
a superpotential, the $A_{\infty}$ structure must have a cyclic structure,
corresponding to rotations of open string disk diagrams.
This was not explicitly addressed in \cite{aspkatz}, and in any event
will be irrelevant for us, as we naturally find $A_{\infty}$ structures
of this form.} so it will give us the same superpotential. 

\subsection{The $A_n$ case}

The simplest case of an obstructed ${\bf P}^1$ is discussed 
in \cite{aspkatz}.  In this example, the normal bundle to a ${\bf P}^1$
in a Calabi-Yau threefold is ${\cal O} \oplus {\cal O}(-2)$, but the
complex structure is not the one inherited from the normal bundle, but rather
is described by the transition functions
\begin{eqnarray*}
w & = & x^{-1} \\
z_1 & = & x^2 y_1 \: + \: x y_2^n \\
z_2 & = & y_2
\end{eqnarray*}
as discussed earlier.

We have already seen that a direct computation of the superpotential
is very difficult, but \cite{aspkatz} 
quickly show that $W = x^{n+1}$, which nicely corresponds to 
a Landau-Ginzburg minimal model superpotential.

More general cases of obstructed curves were not worked out in
\cite{aspkatz}, though their methods certainly apply; we compute them below.

\subsection{The $D_{n+2}$ case}

Next, we will consider the total space $X$ of an obstructed bundle 
over a curve $C\cong\mathbb{P}^1$ with normal bundle 
$\o_{C}(-3)\oplus\o_{C}(1)$. 
In terms of transition function on two open affine charts with 
coordinates  say $(x,y_1,y_2)$ and $(w,z_1,z_2)$  $X$ can be described by
\begin{eqnarray}
z_1&=&x^3y_1+y_2^2+x^2y_2^k\\
z_2&=&x^{-1}y_2\\
w&=&x^{-1}
\end{eqnarray}

 Let $\pi:X \la C$ be the bundle map and denote with $\o(1)=\pi^*\o_C(1)$. 
We will use the methods of \cite{aspkatz} to compute the 
resulting superpotential.

 Thus we need a locally free resolution of the sheaf $\o_C$. 
One such resolution is given by the complex
$$
\xymatrix@1@C=20mm{
\o(-n-5) \ar[r]^{ \left(\begin{smallmatrix}y_2\\-x^3\\-1 \end{smallmatrix}\right)}&
{\begin{matrix}\o(-n-2)\\\oplus\\\o(-n-2)\\\oplus\\\o(-1)\end{matrix} }\ar[r]^{\left(\begin{smallmatrix} x^3&y_2&0\\s'&-y_1&z_1\\-1&0&-y_2\end{smallmatrix}\right)}& {\begin{matrix}\o(-n+1)\\\oplus\\\o(-1)\\\oplus\\\o\end{matrix} }\ar[r]^{\left(\begin{smallmatrix}y_1&y_2&z_1\end{smallmatrix} \right)}& \o,&
}
$$
where 
$$
s'=y_2+x^2y_2^{n-1}.
$$

The maps are given in the first chart (the one given by $(x,y_1,y_2)$). 
By $z_1$ we mean the 
section of $\o$ which in the first chart is given by $x^3y_1+y_2^2+x^2y_2^n$. 
We are considering $y_1$ as a section of $\o(n-1)$ and this can be done since
 \begin{eqnarray}
 x^{-n+1}y_1&=&x^{-n-2}z_1+x^{-n-2}y_2^2+x^{-n}y_2^{n}\\
&=&           w^{n+2}z_1+w^{n+2}z_2^2+z_2^{n}.
\end{eqnarray}
In a similar way $y_2$ as a section of $\o(1)$ over 
the first chart and $s'$ is a section of $\o(n+1)$. Let us also define 
$s= 1+ x^2y_2^{n-2}$. 

 To simplify notation we will name the sheaves of the above resolution 
$\f_i$, $i=0,1,2,3$ so that that the resolution now is given by 
$$
\xymatrix@1@C=10mm{
0\ar[r]&\f_3 \ar[r]&\f_2 \ar[r]&\f_1 \ar[r]&\f_0 \ar[r]&\o_C \ar[r]&0&
}$$

Corresponding to a class in
\begin{displaymath}
C^0(U,Hom^1(\o_C,\o_C))
\end{displaymath}
let $\x$ be the following generator of $\mbox{Ext}^1(\o_C,\o_C)$:
$$
\xymatrix@1@C=15mm{
&\f_3\ar[d]^{\left(\begin{smallmatrix}1\\0\\0\end{smallmatrix}\right)} \ar[r]&
\f_2\ar[d]^{\left(\begin{smallmatrix} 0&1&0\\-s&0&0\\0&0&-1\end{smallmatrix}\right)} \ar[r]&
\f_1 \ar[d]^{\left(\begin{smallmatrix}0&1&0\end{smallmatrix}\right)} \ar[r]&\f_0\\
\f_3\ar[r]&\f_2 \ar[r]&\f_1 \ar[r]&\f_0
}
$$
and  let  $\y=$
$$
\xymatrix@1@C=15mm{
&\f_3\ar[d]^{\left(\begin{smallmatrix}x\\0\\0\end{smallmatrix}\right)} \ar[r]&
\f_2\ar[d]^{\left(\begin{smallmatrix} 0&x&0\\-xs&0&0\\0&0&-x\end{smallmatrix}\right)} \ar[r]&
\f_1 \ar[d]^{\left(\begin{smallmatrix}0&x&0\end{smallmatrix}\right)} \ar[r]&\f_0\\
\f_3\ar[r]&\f_2 \ar[r]&\f_1 \ar[r]&\f_0
}
$$

First of all let us compute $\x \star \x=$
$$
\xymatrix@1@C=15mm{
&&\f_3\ar[d]^{\left(\begin{smallmatrix} 0\\-s\\0\end{smallmatrix}\right)} \ar[r]&\f_2 \ar[d]^{\left(\begin{smallmatrix} -s&0&0\end{smallmatrix}\right)} \ar[r]&\f_1\ar[r]&\f_0\\
\f_3\ar[r]&\f_2 \ar[r]&\f_1 \ar[r]&\f_0
}
$$

 At this point we will simply define some auxiliary elements 
that will be useful in the derivation of the $A_{\infty}$-structure.

 Let $\j=$
$$
\xymatrix@1@C=15mm{
&\f_3\ar[d]^{\left(\begin{smallmatrix}0\\0\\0\end{smallmatrix}\right)} \ar[r]&
\f_2\ar[d]^{\left(\begin{smallmatrix} 0&0&0\\0&0&1\\0&0&0\end{smallmatrix}\right)} \ar[r]&
\f_1 \ar[d]^{\left(\begin{smallmatrix}0&0&1\end{smallmatrix}\right)} \ar[r]&\f_0\\
\f_3\ar[r]&\f_2 \ar[r]&\f_1 \ar[r]&\f_0
}
$$

 Note that $d\j=$
$$
\xymatrix@1@C=15mm{
&&\f_3\ar[d]^{\left(\begin{smallmatrix} 0\\-1\\0\end{smallmatrix}\right)} \ar[r]&\f_2 \ar[d]^{\left(\begin{smallmatrix} -1&0&0\end{smallmatrix}\right)} \ar[r]&\f_1\ar[r]&\f_0\\
\f_3\ar[r]&\f_2 \ar[r]&\f_1 \ar[r]&\f_0
}
$$
and we have the important commutation relation, namely
$$
\j\star\x+\x\star\j=0.
$$

 For $p=0,..n+1$ set $\k_p=$
$$
\xymatrix@1@C=15mm{
&\f_3\ar[d]^{\left(\begin{smallmatrix}0\\0\\0\end{smallmatrix}\right)} \ar[r]&
\f_2\ar[d]^{\left(\begin{smallmatrix} 0&0&0\\x^2y_2^{p}&0&0\\0&0&0\end{smallmatrix}\right)} \ar[r]&
\f_1 \ar[d]^{\left(\begin{smallmatrix}0&0&0\end{smallmatrix}\right)} \ar[r]&\f_0\\
\f_3\ar[r]&\f_2 \ar[r]&\f_1 \ar[r]&\f_0
}
$$
and compute that for $p=1,..,n$ the differential $d\k_{p-1}=:\F_p$  is  
$$
\xymatrix@1@C=15mm{
&&\f_3\ar[d]^{\left(\begin{smallmatrix} 0\\-x^2y_2^p\\0\end{smallmatrix}\right)} \ar[r]&\f_2 
\ar[d]^{\left(\begin{smallmatrix} -x^2y_2^p&0&0\end{smallmatrix}\right)} \ar[r]&\f_1\ar[r]&\f_0\\
\f_3\ar[r]&\f_2 \ar[r]&\f_1 \ar[r]&\f_0
}
$$

 We have the folloing relations:
$$\k_p\star\x+\x\star\k_p=\F_p$$
 $$
\k_j\star\k_i=0
$$

 Observe that 
 $$
 \x\star\x=d(\j+\k_{n-3})
$$

This is enough to compute $m_2$ by using 
 $$
im_2(\x,\x)=i(\x\star\x)+df_2(\x,\x)
$$
and so using what we have computed for $\x\star\x$  we have
$$
im_2(\x,\x)=d(\j+\k_{n-3})+df_2(\x,\x)
$$
So that 
$$
m_2(\x,\x)=0,\qquad f_2(\x,\x)=-(\j+\k_{n-3}).
$$
 Using the next $A_{\infty}$-morphism relation we have 
$$
im_3(\x,\x,\x)=f_2(\x\otimes m_2(\x,\x))-f_2(m_2(\x,\x)\otimes \x)+\x\star f_2(\x,\x)-f_2(\x,\x)\star\x+df_2.
$$
But this is 
$$
im_3(\x,\x,\x)=-\x\star (\j+\k_{n-3})-(\j+\k_{n-3})\star\x+df_3
$$
or
$$
im_3(\x,\x,\x)=-\F_{n-3}+df_3=-d\k_{n-4}+df_3(\x,\x,\x).
$$

So that $m_3(\x,\x,\x)=0$ and $f_3(\x,\x,\x)=\k_{n-4}$.

 Proceeding like this we see that $m_{l}(\x,...,\x)=0$ and 
$f_i=(-1)^{\frac{l(l-1)}{2}}\k_{n-l-1}$ for $2 \le l < n$. Also 
$$
m_{n}(\x,...,\x)=-(-1)^{\frac{n(n-1)}{2}}\F_0.$$
But $\x\star \F_0$ is a generator of $\mbox{Ext}^1(\o_C,\o_C)$. 
So in the superpotential we have a term equal to 
$-(-1)^{\frac{l(l-1)}{2}}x^{n+1}$. 
Notice that $\x\star\x$ is exact and $\y$ is closed so that 
$\x\star\x\star\y$ is also exact. On the other hand 
$\y\star\y\star\y$ is given by 

This implies that  $m_3(\x,\x,\x)=-F_0$ and $f_3(\x,\x,\x)=0$ since  $F_0$ is not exact. Notice that $\x\star m_3(\x,\x,\x)=\x\star F_0=\F_0\star \x$ is 
$$
\xymatrix@1@C=15mm{
&&&\f_3\ar[d]^{x^3+x^5y_2^{n-2}} \ar[r]&\f_2  \ar[r]&\f_1\ar[r]&\f_0\\
\f_3\ar[r]&\f_2 \ar[r]&\f_1 \ar[r]&\f_0&
}
$$
which is the differential of 
$$
\xymatrix@1@C=15mm{
&&\f_3\ar[d]^{\left(\begin{smallmatrix} 0\\0\\0\end{smallmatrix}\right)} \ar[r]&\f_2 \ar[d]^{\left(\begin{smallmatrix} x^5y_2^{n-3}&-1&0\end{smallmatrix}\right)} \ar[r]&\f_1\ar[r]&\f_0\\
\f_3\ar[r]&\f_2 \ar[r]&\f_1 \ar[r]&\f_0
}
$$
 It is easy to check that $\y\star\y\star\x$ generates 
$\mbox{Ext}^3(\o_C,\o_C)$ so that in the superpotential we have a term 
of the type $xy^2$.  One can see immediately that all the higher 
products vanish so that the superpotential is 
$$
W(x,y)=-(-1)^{\frac{n(n-1)}{2}}x^{n+1}+xy^2.
$$

 \subsection{The $E_7$ case.}

Let us examine the $E_7$ case before the $E_6$ and $E_8$ cases,
as it is slightly more complicated, so once we understand
$E_7$, the remaining two cases will be comparatively easy.

The transition functions for the two affine charts 
$(x,y_1,y_2)$ and $(w,z_1,z_2)$ are now 
\begin{eqnarray}
z_1&=&x^3y_1+x y_2^3+x^{-1}y_2^{2}\\
z_2&=&x^{-1}y_2\\
w&=&x^{-1}
\end{eqnarray}

 Proceeding as before we consider a resolution of  $\o_C$, for example 
$$
\xymatrix@1@C=20mm{
\o(-6) \ar[r]^{ \left(\begin{smallmatrix}y_2\\-x^4\\-1 \end{smallmatrix}\right)}&
{\begin{matrix}\o(-5)\\\oplus\\\o(-2)\\\oplus\\\o(-2)\end{matrix} }\ar[r]^{\left(\begin{smallmatrix} x^4&y_2&0\\s'&-y_1&t\\-1&0&-y_2\end{smallmatrix}\right)}& {\begin{matrix}\o(-1)\\\oplus\\\o(-1)\\\oplus\\\o(-1)\end{matrix} }\ar[r]^{\left(\begin{smallmatrix}y_1&y_2&t\end{smallmatrix} \right)}& \o.&
}
$$
where we have defined 
$$
t=x^4y_1+x^2y_2^3+y_2^2,
$$
considered as a  section of $\o(1)$ and 
$$
s'=y_2+x^2y_2^{2},
$$ 
considered as a  section of $\o(4)$. Also $y_1$ and $y_2$ here are 
considered sections of $\o(1)$.
 To simplify notations let as call the sheaves of the above resolution 
$\f_i$ so that we have 

$$
\xymatrix@1@C=10mm{
0\ar[r]&\f_3 \ar[r]&\f_2 \ar[r]&\f_1 \ar[r]&\f_0 \ar[r]&\o_C \ar[r]&0&
}$$

Corresponding to a class in
\begin{displaymath}
C^0(U,Hom^1(\o_C,\o_C))
\end{displaymath}
let $\x$ be the following generator of $\mbox{Ext}^1(\o_C,\o_C)$:
$$
\xymatrix@1@C=15mm{
&\f_3\ar[d]^{\left(\begin{smallmatrix}1\\0\\0\end{smallmatrix}\right)} \ar[r]&
\f_2\ar[d]^{\left(\begin{smallmatrix} 0&1&0\\-s&0&0\\0&0&-1\end{smallmatrix}\right)} \ar[r]&
\f_1 \ar[d]^{\left(\begin{smallmatrix}0&1&0\end{smallmatrix}\right)} \ar[r]&\f_0\\
\f_3\ar[r]&\f_2 \ar[r]&\f_1 \ar[r]&\f_0,
}
$$
where  $s=1+x^2y_2$. Define another generator $\y$ of $\mbox{Ext}^1(\o_C,\o_C)$
$$
\xymatrix@1@C=15mm{
&\f_3\ar[d]^{\left(\begin{smallmatrix}x\\0\\0\end{smallmatrix}\right)} \ar[r]&
\f_2\ar[d]^{\left(\begin{smallmatrix} 0&x&0\\-xs&0&0\\0&0&-x\end{smallmatrix}\right)} \ar[r]&
\f_1 \ar[d]^{\left(\begin{smallmatrix}0&x&0\end{smallmatrix}\right)} \ar[r]&\f_0\\
\f_3\ar[r]&\f_2 \ar[r]&\f_1 \ar[r]&\f_0
}
$$

 Let us compute $\x \star \x=$
$$
\xymatrix@1@C=15mm{
&&\f_3\ar[d]^{\left(\begin{smallmatrix} 0\\-s\\0\end{smallmatrix}\right)} \ar[r]&\f_2 \ar[d]^{\left(\begin{smallmatrix} -s&0&0\end{smallmatrix}\right)} \ar[r]&\f_1\ar[r]&\f_0\\
\f_3\ar[r]&\f_2 \ar[r]&\f_1 \ar[r]&\f_0
}
$$

Let $\j=$
$$
\xymatrix@1@C=15mm{
&\f_3\ar[d]^{\left(\begin{smallmatrix}0\\0\\0\end{smallmatrix}\right)} \ar[r]&
\f_2\ar[d]^{\left(\begin{smallmatrix} 0&0&0\\0&0&1\\0&0&0\end{smallmatrix}\right)} \ar[r]&
\f_1 \ar[d]^{\left(\begin{smallmatrix}0&0&1\end{smallmatrix}\right)} \ar[r]&\f_0\\
\f_3\ar[r]&\f_2 \ar[r]&\f_1 \ar[r]&\f_0
}
$$

 Note that $d\j=$
$$
\xymatrix@1@C=15mm{
&&\f_3\ar[d]^{\left(\begin{smallmatrix} 0\\-1\\0\end{smallmatrix}\right)} \ar[r]&\f_2 \ar[d]^{\left(\begin{smallmatrix} -1&0&0\end{smallmatrix}\right)} \ar[r]&\f_1\ar[r]&\f_0\\
\f_3\ar[r]&\f_2 \ar[r]&\f_1 \ar[r]&\f_0
}
$$

and we have this important commutation relation, namely
$$
\j\star\x+\x\star\j=0.
$$
For $p=0,1$ we set $\k_p=$
$$
\xymatrix@1@C=15mm{
&\f_3\ar[d]^{\left(\begin{smallmatrix}0\\0\\0\end{smallmatrix}\right)} \ar[r]&
\f_2\ar[d]^{\left(\begin{smallmatrix} 0&0&0\\x^2y_2^{p}&0&0\\0&0&0\end{smallmatrix}\right)} \ar[r]&
\f_1 \ar[d]^{\left(\begin{smallmatrix}0&0&0\end{smallmatrix}\right)} \ar[r]&\f_0\\
\f_3\ar[r]&\f_2 \ar[r]&\f_1 \ar[r]&\f_0
}
$$
and compute $d\k_{p-1}=:\F_p$ to be 
$$
\xymatrix@1@C=15mm{
&&\f_3\ar[d]^{\left(\begin{smallmatrix} 0\\-x^2y_2^p\\0\end{smallmatrix}\right)} \ar[r]&\f_2 
\ar[d]^{\left(\begin{smallmatrix} -x^2y_2^p&0&0\end{smallmatrix}\right)} \ar[r]&\f_1\ar[r]&\f_0\\
\f_3\ar[r]&\f_2 \ar[r]&\f_1 \ar[r]&\f_0
}
$$

 Let us write down the last few relations that we need to compute the 
$A_{\infty}$-structure.

$$\k_p\star\x+\x\star\k_p=\F_p$$
 $$
\k_j\star\k_i=0
$$

 Now, first we have 
 $$
 \x\star\x==d\j+F_1=d(\j+\k_0)
$$
$$
im_2(\x,\x)=i(\x\star\x)+df_2(\x,\x)
$$
and using \cite{aspkatz} we have
$$
im_2(\x,\x)=d(\j+\k_0)+df_2(\x,\x)
$$
So that 
$$
m_2(\x,\x)=0,\qquad f_2(\x,\x)=-(\j+\k_0).
$$
 Using the next $A_{\infty}$-morphism relation we have 
$$
im_3(\x,\x,\x)=f_2(\x\otimes m_2(\x,\x))-f_2(m_2(\x,\x)\otimes \x)+\x\star f_2(\x,\x)-f_2(\x,\x)\star\x+df_2
$$

But this is 
$$
im_3(\x,\x,\x)=-\x\star (\j+\k_0)-(\j+\k_0)\star\x+df_3
$$
or
$$
im_3(\x,\x,\x)=-\F_0+df_3.
$$

This implies that  $m_3(\x,\x,\x)=-F_0$ and $f_3(\x,\x,\x)=0$ 
since  $F_0$ is not exact. 
Notice that $\x\star m_3(\x,\x,\x)=\x\star F_0=\F_0\star \x$ is 
$$
\xymatrix@1@C=15mm{
&&&\f_3\ar[d]^{x^2} \ar[r]&\f_2  \ar[r]&\f_1\ar[r]&\f_0\\
\f_3\ar[r]&\f_2 \ar[r]&\f_1 \ar[r]&\f_0&
}
$$
which is the differential of 
$$
\xymatrix@1@C=15mm{
&&\f_3\ar[d]^{\left(\begin{smallmatrix} 0\\0\\0\end{smallmatrix}\right)} \ar[r]&\f_2 \ar[d]^{\left(\begin{smallmatrix} 0&0&-x^2\end{smallmatrix}\right)} \ar[r]&\f_1\ar[r]&\f_0\\
\f_3\ar[r]&\f_2 \ar[r]&\f_1 \ar[r]&\f_0
}
$$
 It is easy to check that $\y\star F_0$ generates 
$\mbox{Ext}^3(\o_C,\o_C)$ so that in the superpotential we have a 
term of the type $yx^3$. 
Also if we compute $\y \star \y\star\y$ we obtain 
$$
\xymatrix@1@C=15mm{
&&&\f_3\ar[d]^{x^3s} \ar[r]&\f_2  \ar[r]&\f_1\ar[r]&\f_0\\
\f_3\ar[r]&\f_2 \ar[r]&\f_1 \ar[r]&\f_0&
}
$$
which also generates so we have a term of the type $y^3$. 
One can see immediately that all the higher products vanish.

\subsection{The $E_6$ case.}

 We procced with the $E_6$ case.
The change of coordinates is given by 
\begin{eqnarray}
z_1&=&x^3y_1+x^2 y_2^3+x^{-1}y_2^{2}\\
z_2&=&x^{-1}y_2\\
w&=&x^{-1}
\end{eqnarray}

The resolution of $\o_C$ is:
$$
\xymatrix@1@C=32mm{
\o(-7) \ar[r]^{ \left(\begin{smallmatrix}y_2\\-x^4\\-1 \end{smallmatrix}\right)}
&
{\begin{matrix}\o(-6)\\\oplus\\\o(-2)\\\oplus\\\o(-2)\end{matrix} }\ar[r]^{
\left(
\begin{smallmatrix} x^4&y_2&0\\s'&-y_1&t\\-1&0&-y_2\end{smallmatrix}\right)}& {
\begin{matrix}\o(-2)\\\oplus\\\o(-1)\\\oplus\\\o(-1)\end{matrix} }\ar[r]^{\left(
\begin{smallmatrix}y_1&y_2&t\end{smallmatrix} \right)}& \o.&
}
$$
where we have
$$
t=x^4y_1+x^3y_2^3+y_2^2,
$$
a section of $\o(1)$ and 
$$
s'=y_2+x^3y_2^{2},
$$ 
a section of $\o(4)$.  Also, $y_1$ is a section of $\o(2)$ and $y_2$ is
a section of
$\o(1)$.
 To simplify notation let us call the sheaves of the above resolution 
$\f_i$ so that we have 
$$
\xymatrix@1@C=10mm{
0\ar[r]&\f_3 \ar[r]&\f_2 \ar[r]&\f_1 \ar[r]&\f_0 \ar[r]&\o_C \ar[r]&0&
}$$

Corresponding to a class in
\begin{displaymath}
C^0(U,Hom^1(\o_C,\o_C))
\end{displaymath}
let  $\x$  be the following generator of $\mbox{Ext}^1(\o_C,\o_C)$:
$$
\xymatrix@1@C=15mm{
&\f_3\ar[d]^{\left(\begin{smallmatrix}1\\0\\0\end{smallmatrix}\right)} \ar[r]&
\f_2\ar[d]^{\left(\begin{smallmatrix} 0&1&0\\-s&0&0\\0&0&-1\end{smallmatrix}
\right)}
\ar[r]&
\f_1 \ar[d]^{\left(\begin{smallmatrix}0&1&0\end{smallmatrix}\right)} \ar[r]&\f_0
\\
\f_3\ar[r]&\f_2 \ar[r]&\f_1 \ar[r]&\f_0,
}
$$
where  $s=1+x^3y_2^2$. Define another generator $\y$ of $\mbox{Ext}^1(\o_C,\o_C)$
$$
\xymatrix@1@C=15mm{
&\f_3\ar[d]^{\left(\begin{smallmatrix}x\\0\\0\end{smallmatrix}\right)} \ar[r]&
\f_2\ar[d]^{\left(\begin{smallmatrix} 0&x&0\\-xs&0&0\\0&0&-x\end{smallmatrix}
\right)} \ar[r]&
\f_1 \ar[d]^{\left(\begin{smallmatrix}0&x&0\end{smallmatrix}\right)} \ar[r]&\f_0
\\
\f_3\ar[r]&\f_2 \ar[r]&\f_1 \ar[r]&\f_0
}
$$

 Let us compute $\x \star \x=$
$$
\xymatrix@1@C=15mm{
&&\f_3\ar[d]^{\left(\begin{smallmatrix} 0\\-s\\0\end{smallmatrix}\right)} \ar[r]
&\f_2 \ar[d]^{\left(\begin{smallmatrix} -s&0&0\end{smallmatrix}\right)} \ar[r]&
\f_1\ar[r]&\f_0\\
\f_3\ar[r]&\f_2 \ar[r]&\f_1 \ar[r]&\f_0
}
$$

Let $\j=$
$$
\xymatrix@1@C=15mm{
&\f_3\ar[d]^{\left(\begin{smallmatrix}0\\0\\0\end{smallmatrix}\right)} \ar[r]&
\f_2\ar[d]^{\left(\begin{smallmatrix} 0&0&0\\0&0&1\\0&0&0\end{smallmatrix}\right
)} \ar[r]&
\f_1 \ar[d]^{\left(\begin{smallmatrix}0&0&1\end{smallmatrix}\right)} \ar[r]&\f_0
\\
\f_3\ar[r]&\f_2 \ar[r]&\f_1 \ar[r]&\f_0
}
$$

 Note that $d\j=$
$$
\xymatrix@1@C=15mm{
&&\f_3\ar[d]^{\left(\begin{smallmatrix} 0\\-1\\0\end{smallmatrix}\right)} \ar[r]
&\f_2 \ar[d]^{\left(\begin{smallmatrix} -1&0&0\end{smallmatrix}\right)} \ar[r]&\
\f_1\ar[r]&\f_0\\
\f_3\ar[r]&\f_2 \ar[r]&\f_1 \ar[r]&\f_0
}
$$

and we have this important commutation relation, namely
$$
\j\star\x+\x\star\j=0.
$$
For $p=0,1$ we set $\k_p=$
$$
\xymatrix@1@C=15mm{
&\f_3\ar[d]^{\left(\begin{smallmatrix}0\\0\\0\end{smallmatrix}\right)} \ar[r]&
\f_2\ar[d]^{\left(\begin{smallmatrix} 0&0&0\\x^3y_2^{p}&0&0\\0&0&0
\end{smallmatrix}
\right)} \ar[r]&
\f_1 \ar[d]^{\left(\begin{smallmatrix}0&0&0\end{smallmatrix}\right)} \ar[r]&\f_0
\\
\f_3\ar[r]&\f_2 \ar[r]&\f_1 \ar[r]&\f_0
}
$$
and compute $d\k_{p-1}=:\F_p$ to be 
$$
\xymatrix@1@C=15mm{
&&\f_3\ar[d]^{\left(\begin{smallmatrix} 0\\-x^3y_2^p\\0\end{smallmatrix}\right)}
 \ar[r]&\f_2 
\ar[d]^{\left(\begin{smallmatrix} -x^3y_2^p&0&0\end{smallmatrix}\right)} \ar[r]&
\f_1\ar[r]&\f_0\\
\f_3\ar[r]&\f_2 \ar[r]&\f_1 \ar[r]&\f_0
}
$$

 Let us write down the last few relations that we need to compute the 
$A_{\infty}$-structure.

$$\k_p\star\x+\x\star\k_p=\F_p$$
 $$
\k_j\star\k_i=0
$$

 Now, first we have 
 $$
 \x\star\x=d\j+F_1=d(\j+\k_0)
$$
$$
im_2(\x,\x)=i(\x\star\x)+df_2(\x,\x)
$$
and thus
$$
im_2(\x,\x)=d(\j+\k_0)+df_2(\x,\x)
$$
So that 
$$
m_2(\x,\x)=0,\qquad f_2(\x,\x)=-(\j+\k_0).
$$
 Using the next $A_{\infty}$-morphism relation we have 
$$
im_3(\x,\x,\x)=f_2(\x\otimes m_2(\x,\x))-f_2(m_2(\x,\x)\otimes \x)+\x\star f_2(\
x,\x)-f_2(\x,\x)\star\x+df_2
$$

But this is 
$$
im_3(\x,\x,\x)=-\x\star (\j+\k_0)-(\j+\k_0)\star\x+df_3
$$
or
$$
im_3(\x,\x,\x)=-\F_0+df_3.
$$

This implies that  $m_3(\x,\x,\x)=-\F_0$ and $f_3(\x,\x,\x)=0$ since  $\F_0$ is 
not exact.  Notice that $\x\star m_3(\x,\x,\x)=\x\star \F_0=\F_0\star \x$ is 
$$
\xymatrix@1@C=15mm{
&&&\f_3\ar[d]^{x^3} \ar[r]&\f_2  \ar[r]&\f_ 1\ar[r]&\f_0\\
\f_3\ar[r]&\f_2 \ar[r]&\f_1 \ar[r]&\f_0&
}
$$
which generates $\mbox{Ext}^3(\o_C,\o_C)$  so in the superpotential we have $x^4$. 
 It is easy to check that $\y \star \y\star\y$ also generates 
$\mbox{Ext}^3(\o_C,\o_C)$
 so that in the superpotential we have a term of the type $y^3$. 
All the higher 
products vanish.  So the superpotential is 
$$
W=x^4+y^3
$$

\subsection{The $E_8$ case.}

The transition functions for the two affine charts 
$(x,y_1,y_2)$ and $(w,z_1,z_2)$
are now 
\begin{eqnarray}
z_1&=&x^3y_1+x^2 y_2^4+x^{-1}y_2^{2}\\
z_2&=&x^{-1}y_2\\
w&=&x^{-1}
\end{eqnarray}

 Proceeding as before we consider a resolution of  $\o_C$, for example 
$$
\xymatrix@1@C=20mm{
\o(-8) \ar[r]^{ \left(\begin{smallmatrix}y_2\\-x^4\\-1 \end{smallmatrix}\right)}
&
{\begin{matrix}\o(-7)\\\oplus\\\o(-2)\\\oplus\\\o(-2)\end{matrix} }\ar[r]^{\left
(\begin{smallmatrix} x^4&y_2&0\\s'&-y_1&t\\-1&0&-y_2\end{smallmatrix}\right)}& {
\begin{matrix}\o(-3)\\\oplus\\\o(-1)\\\oplus\\\o(-1)\end{matrix} }\ar[r]^{\left(
\begin{smallmatrix}y_1&y_2&t\end{smallmatrix} \right)}& \o.&
}
$$
where we have defined 
$$
t=x^4y_1+x^3y_2^4+y_2^2,
$$
considered as a  section of $\o(1)$ and 
$$
s'=y_2+x^3y_2^{3},
$$ 
considered as a  section of $\o(4)$. Also $y_1$ is a section of $\o(2)$  and 
$y_2$
is a section of $\o(1)$.
 To simplify notation let us call the sheaves of the above resolution 
$\f_i$ so that we have 
$$
\xymatrix@1@C=10mm{
0\ar[r]&\f_3 \ar[r]&\f_2 \ar[r]&\f_1 \ar[r]&\f_0 \ar[r]&\o_C \ar[r]&0&
}$$

Corresponding to a class in
\begin{displaymath}
C^0(U,Hom^1(\o_C,\o_C))
\end{displaymath}
let $\x$  be the following generator of $\mbox{Ext}^1(\o_C,\o_C)$: 
$$
\xymatrix@1@C=15mm{
&\f_3\ar[d]^{\left(\begin{smallmatrix}1\\0\\0\end{smallmatrix}\right)} \ar[r]&
\f_2\ar[d]^{\left(\begin{smallmatrix} 0&1&0\\-s&0&0\\0&0&-1\end{smallmatrix}
\right)}
 \ar[r]&
\f_1 \ar[d]^{\left(\begin{smallmatrix}0&1&0\end{smallmatrix}\right)} \ar[r]&\f_0
\\
\f_3\ar[r]&\f_2 \ar[r]&\f_1 \ar[r]&\f_0,
}
$$
where  $s=1+x^3y_2^2$.  Define another generator $\y$ of $\mbox{Ext}^1(\o_C,\o_C)$
$$
\xymatrix@1@C=15mm{
&\f_3\ar[d]^{\left(\begin{smallmatrix}x\\0\\0\end{smallmatrix}\right)} \ar[r]&
\f_2\ar[d]^{\left(\begin{smallmatrix} 0&x&0\\-xs&0&0\\0&0&-x\end{smallmatrix}
\right)} \ar[r]&
\f_1 \ar[d]^{\left(\begin{smallmatrix}0&x&0\end{smallmatrix}\right)} \ar[r]&\f_0
\\
\f_3\ar[r]&\f_2 \ar[r]&\f_1 \ar[r]&\f_0
}
$$

 Let us compute $\x \star \x=$
$$
\xymatrix@1@C=15mm{
&&\f_3\ar[d]^{\left(\begin{smallmatrix} 0\\-s\\0\end{smallmatrix}\right)} \ar[r]
&\f_2 \ar[d]^{\left(\begin{smallmatrix} -s&0&0\end{smallmatrix}\right)} \ar[r]&\
\f_1\ar[r]&\f_0\\
\f_3\ar[r]&\f_2 \ar[r]&\f_1 \ar[r]&\f_0
}
$$

Let $\j=$
$$
\xymatrix@1@C=15mm{
&\f_3\ar[d]^{\left(\begin{smallmatrix}0\\0\\0\end{smallmatrix}\right)} \ar[r]&
\f_2\ar[d]^{\left(\begin{smallmatrix} 0&0&0\\0&0&1\\0&0&0\end{smallmatrix}\right
)} \ar[r]&
\f_1 \ar[d]^{\left(\begin{smallmatrix}0&0&1\end{smallmatrix}\right)} \ar[r]&\f_0
\\
\f_3\ar[r]&\f_2 \ar[r]&\f_1 \ar[r]&\f_0
}
$$

 Note that $d\j=$
$$
\xymatrix@1@C=15mm{
&&\f_3\ar[d]^{\left(\begin{smallmatrix} 0\\-1\\0\end{smallmatrix}\right)} \ar[r]
&\f_2 \ar[d]^{\left(\begin{smallmatrix} -1&0&0\end{smallmatrix}\right)} \ar[r]&\
\f_1\ar[r]&\f_0\\
\f_3\ar[r]&\f_2 \ar[r]&\f_1 \ar[r]&\f_0
}
$$
and we have the important commutation relation, namely
$$
\j\star\x+\x\star\j=0.
$$
For $p=0,1,2$ we set $\k_p=$
$$
\xymatrix@1@C=15mm{
&\f_3\ar[d]^{\left(\begin{smallmatrix}0\\0\\0\end{smallmatrix}\right)} \ar[r]&
\f_2\ar[d]^{\left(\begin{smallmatrix} 0&0&0\\x^3y_2^{p}&0&0\\0&0&0
\end{smallmatrix}
\right)} \ar[r]&
\f_1 \ar[d]^{\left(\begin{smallmatrix}0&0&0\end{smallmatrix}\right)} \ar[r]&\f_0
\\
\f_3\ar[r]&\f_2 \ar[r]&\f_1 \ar[r]&\f_0
}
$$
and compute $d\k_{p-1}=:\F_p$ to be 
$$
\xymatrix@1@C=15mm{
&&\f_3\ar[d]^{\left(\begin{smallmatrix} 0\\-x^3y_2^p\\0\end{smallmatrix}\right)}
 \ar[r]&\f_2 
\ar[d]^{\left(\begin{smallmatrix} -x^3y_2^p&0&0\end{smallmatrix}\right)} \ar[r]&
\f_1\ar[r]&\f_0\\
\f_3\ar[r]&\f_2 \ar[r]&\f_1 \ar[r]&\f_0
}
$$

 Let us write down the last few relations that we need to compute 
the $A_{\infty}$
-structure.
$$\k_p\star\x+\x\star\k_p=\F_p$$
 $$
\k_j\star\k_i=0
$$

 Now, first we have 
 $$
 \x\star\x=d\j+F_2=d(\j+\k_1)
$$
$$
im_2(\x,\x)=i(\x\star\x)+df_2(\x,\x)
$$
and so
$$
im_2(\x,\x)=d(\j+\k_1)+df_2(\x,\x)
$$
So that 
$$
m_2(\x,\x)=0,\qquad f_2(\x,\x)=-(\j+\k_1).
$$
 Using the next $A_{\infty}$-morphism relation we have 
$$
im_3(\x,\x,\x)=f_2(\x\otimes m_2(\x,\x))-f_2(m_2(\x,\x)\otimes \x)+\x\star f_2(\
x,\x)-f_2(\x,\x)\star\x+df_2
$$

But this is 
$$
im_3(\x,\x,\x)=-\x\star (\j+\k_1)-(\j+\k_1)\star\x+df_3
$$
or
$$
im_3(\x,\x,\x)=-\F_1+df_3=-d\k_0+df_3.
$$

This implies that  $m_3(\x,\x,\x)=0$ and $f_3(\x,\x,\x)=\k_0$. Proceedaing like 
before we find that $m_4(\x,\x,\x,\x)=\F_0$ and all the higher products in $\x$ 
vanish.  Since $\x\star \F_0$ generates $\mbox{Ext}^3(\o_C,\o_C)$, in the 
superpotential
we have a term  $x^5$.
 In a similar way we find a term $y^3$ so that finaly the superpotential is
$$
W=x^5+y^3.
$$

\section{Conclusions}

In this short note we have derived superpotentials from D-branes wrapped
on obstructed curves, finding a relationship to 
Landau-Ginzburg presentations of minimal models.

\section{Acknowledgements}

We would like to thank C.~Curto and especially
E.~Sharpe for numerous conversations.

\end{document}